\documentstyle[epsf,seceq]{ptptex}

\newcommand{\bfsigma}{\mbox{\boldmath$\sigma$}}
\newcommand{\bfnabla}{\mbox{\boldmath$\nabla$}}

\title{
Meson Condensations in High-Density Matter 
}

\author{
Takumi {\sc Muto}\footnote{E-mail address: muto@pf.it-chiba.ac.jp} 
}

\inst{
Department of Physics, Chiba Institute of Technology, 2-1-1 Shibazono, 
Narashino, Chiba 275-0023, Japan}

\recdate{
\today}

\abst{Recent studies on meson condensations (pions and kaons) in high-density hadronic matter are reviewed. After summarizing onset mechanisms of pion and kaon condensations, we discuss implications for neutron star phenomena such as rapid cooling through neutrino emission, static and dynamic properties of neutron stars. Recent studies on coexistence problem of pion and kaon condensations are briefly introduced. Finally, relevance of meson condensations in hadronic matter to those in color superconductivity are briefly mentioned. 
}

\begin{document}

\maketitle

\section{Introduction}

Meson condensations (pions and kaons) have been extensively investigated as new states of high-density hadronic matter, which may be realized in neutron stars or may be produced in high energy heavy-ion collisions.\cite{s72,m78,bc79,kmttt93,kn86,t95,lbmr95,l96,pbpelk97} They are characterized as macroscopic appearance of Nambu-Goldstone (NG)  bosons. Different from usual Bose-Einstein condensation in condensed matter physics, meson condensed states are strongly interacting systems of mesons and baryons.  
The critical densities and equations of state (EOS) of pion and kaon condensations have been widely discussed with reference to astrophysical phenomena and nuclear experiments. 

Condensations of the NG bosons are formulated on the basis of chiral symmetry and its spontaneous breakdown. For instance, a charged pion-condensed state $|\pi\rangle$ is given by a chiral rotation from  a normal meson vacuum $| 0 \rangle$  such that $|\pi \rangle=\hat U (\theta, \chi)  |0 \rangle$ with a unitary operator $\hat U (\theta, \chi) $ generating a meson-condensed state, \cite{bc79,kmttt93}
\begin{equation}
\hat U (\theta, \chi)=\exp (i\int d^3 r \chi V_3^0 ) \exp (iQ_1^5\theta) \ . 
\label{eq:u}
\end{equation}
In Eq.~(\ref{eq:u}),  $\chi$ and $\theta$ (the chiral angle) are quantities characterizing a type of condensation, and $V_3^\mu$ and $Q_1^5$ the vector current and the axial charge, respectively.  With the help of current algebra and PCAC (partial conservation of axial-vector current), the classical pion field is given by
\begin{equation}
\langle \pi\rangle=\langle\pi|\hat \pi|\pi\rangle=\langle 0| \hat U^\dagger (\theta, \chi) \hat\pi \hat U(\theta, \chi) |0\rangle=
\frac{f}{\sqrt{2}}\sin\theta e^{i\chi} \ , 
\label{eq:order-parameter}
\end{equation}
where $f$ is the meson decay constant. The classical field [ Eq.~(\ref{eq:order-parameter}) ]  specifies an order parameter of condensation. 

At high baryon number density or high temperature, quark degrees of freedom become apparent through chiral and deconfinement phase transitions. Recently, rich phase structure of quark matter has revealed itself reflecting the spin and color-flavor dependence of quark-gluon dynamics, just as hadronic matter reflecting the state dependence of nuclear interactions. In particular, color superconductivity in quark matter such as 2SC, color-flavor locked (CFL) phase etc. have been widely investigated.\cite{a03} Furthermore, condensations of the NG boson modes in CFL phase have been suggested by several authors.\cite{bs02,kr02} 

In order to clarify the whole QCD diagram, an approach from hadronic phase is necessary as well as that from quark phase. 
These quark and hadronic phases may be mutually connected through realization of  meson condensations whose dynamics is controlled by the underlying chiral symmetry. 
In this paper, we give a complementary overview of meson condensations from the point of view of hadronic matter, toward unified understanding of phase structure in high-density QCD. 

\section{Onset mechanisms of meson condensations}

\subsection{$P$-wave pion condensation}
\label{subsec:pion}

The primary part of the $\pi N$ interaction Lagrangian is given by in the lowest order of chiral expansion as  
\begin{equation}
{\cal L}_{\rm int}^{\pi N}=-\tilde f \bar\psi\gamma^\mu\gamma_5\tau_a\psi\partial_\mu\pi^a -\frac{1}{4f^2}\epsilon_{abc}\bar\psi\gamma^\mu \tau^c\psi\pi^a\partial_\mu\pi^b \ , 
\label{eq:pin}
\end{equation}
where $\tilde f\equiv f_{\pi N N}/m_\pi $ with $f_{\pi NN}$($\simeq$1)  being the $\pi N$ coupling constant and $m_\pi$ the pion mass, 
$\psi$ is the nucleon isodoublet, and $a$, $b$, $c$ denote the isospin indices.

The first and second terms on the r.h.s. of Eq.~(\ref{eq:pin}) represent the pseudo-vector coupling and vector coupling, respectively. 
In the nonrelativistic limit, the Lagrangian reduces to 
\begin{eqnarray}
{\cal L}_{\rm int}^{\pi N}&\rightarrow& -\sqrt{2}\tilde f  (\psi^\dagger \bfsigma\tau_-\psi\cdot\bfnabla\pi^- +\psi^\dagger \bfsigma\tau_+\psi\cdot \bfnabla\pi^+) 
-\tilde f \psi^\dagger \bfsigma\tau_3\psi \cdot\bfnabla\pi^0 \cr
&+&\frac{i}{4f^2}\Big\lbrack \sqrt{2}\Big\lbrace\psi^\dagger\tau_+\psi(\pi^0\partial_t \pi^+-\pi^+\partial_t \pi^0)-\psi^\dagger\tau_-\psi(\pi^0\partial_t \pi^- -\pi^-\partial_t \pi^0)
\Big\rbrace \cr
&-&\psi^\dagger\tau_3\psi(\pi^-\partial_t\pi^+-\pi^+\partial_t\pi^-)\Big\rbrack \ , 
\label{eq:nonrelapin}
\end{eqnarray}
where the $\bfsigma\cdot\bfnabla$ couplings give the $p$-wave $\pi$-$N$ attractive interaction leading to the main driving force of pion condensation. 
The remaining terms in Eq.~(\ref{eq:nonrelapin}) give the $s$-wave vector interaction. 

On the other hand, the $s$-wave scalar interaction, the $\pi N$ sigma term $\Sigma_{\pi N}$, comes from the next to leading order of the chiral expansion, and is given as $\displaystyle\Sigma_{\pi N}=\frac{1}{2}(m_u+m_d)\langle N|\bar uu+\bar dd | N\rangle$ with $m_u$ and $m_d$ being the quark masses representing the explicit chiral symmetry breaking. $\Sigma_{\pi N}$ is related to the quark condensate in the nucleon. The value of $\Sigma_{\pi N}$ is empirically estimated to be $\sim$ 45 MeV.\cite{gls91}

As for charged pion ($\pi^c$) condensation, the $s$-wave vector interaction, which corresponds to the Tomozawa-Weinberg term related to the $\pi N$ forward scattering amplitude, is  proportional to the nucleon isospin density $\psi^\dagger\tau_3\psi$ in Eq.~(\ref{eq:nonrelapin}). The $s$-wave $\pi^- p$ vector interaction has an attractive contribution to the pion self-energy, while the $\pi^- n$ vector interaction is repulsive, so that the net vector interaction works repulsively in neutron-star matter. 
At certain density, the $p$-wave $\pi N$ attractive interactions overcome the pion mass and  repulsive $s$-wave vector interactions. It is to be noted that the $s$-wave $\pi N$ interactions given by $\Sigma_{\pi N}$ is relatively weak, which is in contrast to the case of kaon condensation. 

The dispersion relations for $\pi^c$ in neutron-star matter show that there appears a spin-isospin sound mode with a $\pi^+$ quantum number ($\pi_s^+$). At a critical density $\rho_{\rm B}=\rho_{\rm B}^c$, the $\pi^-$ and $\pi_s^+$ branches merge with a pion momentum ${\bf p_\pi}={\bf p}_\pi^c$. The critical density $\rho_{\rm B}^c$ and the momentum ${\bf p}_\pi^c$ are determined from a double-pole condition for the pion inverse propagator, 
\begin{equation}
D^{-1}_\pi (\omega, {\bf p}_\pi; \rho_{\rm B})=0 \ , \partial D^{-1}_\pi( \omega, {\bf p}_\pi; \rho_{\rm B})/\partial \omega=0 \ ,  
\label{eq:double-pole}
\end{equation}
together with $\partial D_\pi^{-1}( \omega, {\bf p}_\pi; \rho_{\rm B})/\partial |{\bf p}_\pi|=0$. At this critical point, the $\pi^-$-$\pi_s^+$ pairs are created with no cost of energy, which implies that the system becomes unstable with respect to condensation of the  $\pi^-$-$\pi_s^+$ pairs.\cite{m78,bc79} \ The critical density $\rho_{\rm B}^c (\pi^c$)  is estimated to be around $2\rho_0$ with $\rho_0$ ($\simeq$0.16 fm$^{-3}$)  being the standard nuclear saturation density. In the condensed phase, due to the isospin-flip through the $p$-wave $\pi N$ interaction, the baryon exists as a quasiparticle $\eta$, which is superposition of the neutron and proton states, i.e., $|\eta_s\rangle=\cos\phi|n_s\rangle\pm i\sin\phi |p_s\rangle$ for $s$=$\pm $1/2, where $s$ denotes the spin state.  

\subsection{Alternating Layer Spin (ALS) structure accompanying $\pi^0$ condensation}
\label{subsec:als}
\vspace{-0.5cm}

\noindent\begin{figure}
\noindent\begin{minipage}[l]{0.5\textwidth} 
\epsfxsize=\textwidth
\centerline{\epsfbox{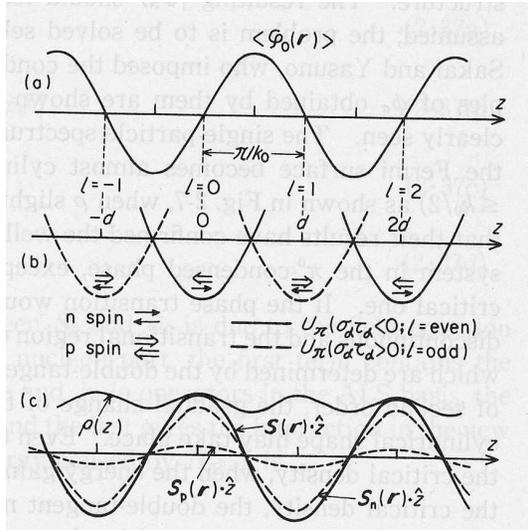}}
\caption{The ALS structure under $\pi^0$ condensation, taken from Ref.~ 15). }
\label{fig1}
\end{minipage}
\begin{minipage}[r]{0.50\textwidth}
\ \ For the neutral pion ($\pi^0$) condensation, the critical density is given by the condition that the excitation energy $\omega$ for the spin-isospin sound mode vanishes : 
$D^{-1}_{\pi^0} (\omega=0, {\bf p}_\pi; \rho_{\rm B})=0 $, together with $\partial D_{\pi^0}^{-1}( \omega, {\bf p}_\pi; \rho_{\rm B})/\partial |{\bf p}_\pi|=0$ at $\omega=0$. 
Since the chemical potential of the $\pi^0$ is zero, the classical $\pi^0$ field can be taken as a standing wave type such that $\langle\varphi_0\rangle=A\sin |{\bf p}_\pi| z$, where 
$A$ is the amplitude of $\pi^0$ condensation. We take the direction of the $\pi^0$ momentum as the $z$ axis.  

 \ \ For the fully developed phase of $\pi^0$ condensation beyond the critical density, Takatsuka et al. presented a model for baryonic configuration in the existence of 
\end{minipage}
\end{figure}~
\vspace{-0.5cm}

\noindent $\pi^0$ condensation as shown in Fig.~\ref{fig1}:\cite{tttt78,ttt93} Under the potential $U_\pi$ generated by $\pi^0$ condensation  ($U_\pi=-\tilde f\tau_3\bfsigma\cdot{\mbox{\boldmath$\nabla$}}\langle\varphi_0\rangle$),  baryons are localized along one dimensional ($z$) direction in a specific spin-isospin order,  and behave as a Fermi gas in the remaining two-dimensional layer. The spin-isospin of baryons changes alternately layer by layer.
Such a baryonic configuration, called the Alternating Layer Spin (ALS) structure, is determined selfconsistently with the classical $\pi^0$ field. The ALS structure gives the most favorable  baryonic configuration, stemming from the one-pion-exchange tensor force. The critical density $\rho_{\rm B}^c$ ({\rm ALS}) is estimated to be $\sim 3 \rho_0$. 
The ALS structure provides us with a new type of quantum solid  
for high-density hadronic matter just like liquid crystals (smectic A phase) in solid state physics. The quantum solid brought about by the ALS structure gives important aspects in high-density hadronic matter in relation to neutron star phenomena, in particular, pulsar glitches (see \ref{subsec:glitch}).  

Subsequently, a possible coexistence of $\pi^0$ and $\pi^c$ condensations has been considered.\cite{tt81} In the combined condensation of $\pi^0$ and $\pi^c$,  the ALS structure develops in the $z$ direction accompanying $\pi^0$ condensation with a standing wave type, and in the remaining two-dimensional layer, $\pi^c$ condensation exists with a plane wave type. 
 The  combined $\pi^0$-$\pi^c$ condensation is energetically  most favorable, and takes over the characteristic features of both $\pi^0$ and $\pi^c$ condensations. 
 
\subsection{Realistic effects on pion condensation}
\label{subsec:realistic}

Pions are easily excited in nuclear matter, and they couple to particle-hole states. The critical density and EOS of pion condensation depend sensitively on the medium effects in the pion channel. Phenomenologically, the following competing effects have been considered in a realistic approach of pion condensation. One is the excitation of the isobar $\Delta $ (1232), which gives strong attraction through $p$-wave $\pi N\Delta$ coupling strength $\displaystyle f_{\pi N\Delta}=\sqrt{\frac{72}{25}}f_{\pi NN} $ in the quark model and is favorable for pion condensation. 
Another is short-range correlation between baryons, which is simulated by the Landau-Migdal (LM) parameter in the spin-isospin channel and is unfavorable for pion condensation. 
The LM parameter is denoted as $g'_{NN}$ for correlations between the $NN^{-1}$ and $NN^{-1}$ particle-hole pairs (`$-1$' denotes a hole state), $g'_{N\Delta}$ for $NN^{-1}$ and $\Delta N^{-1}$ particle-hole pairs, and  $g'_{\Delta\Delta}$ for $\Delta N^{-1}$ and $\Delta N^{-1}$ particle-hole pairs. The value of the LM parameter was evaluated phenomenologically or based on the reaction matrix calculations. However, in most of the calculations in nuclear matter, the universality was assumed, i. e., $g'\equiv g'_{NN}=g'_{N\Delta}=g'_{\Delta\Delta}$, and a rather large value, $g'$=0.7$-$0.9, was obtained. \cite{ew88}
 
 Recently, new information on the LM parameters has been obtained from the experiment of the giant Gamow-Teller (GT) resonances, which are relevant to the spin-isospin sound mode of the kinematical region $\omega\sim 0$, $|{\bf p}_\pi|\sim 0$.\cite{w97}  From the analysis of the GT sum rule for the reaction, $^{90}$Zr ($p,n$) $^{90}$Nb at 295 MeV, the quenching factor has been estimated to be $Q=90\pm 5$\% , which leads to constraint on the Landau-Migdal parameters as\cite{ss99}
\begin{equation}
g'_{N\Delta}=0.18+0.05g'_{\Delta\Delta} \ \ {\rm and} \ \ g'_{NN}\simeq 0.59 \ \ {\rm for}  \ \ g'_{\Delta\Delta} <1 \ . 
\label{eq:gprime}
\end{equation}
This result shows that the universality is largely broken. 
By the use of the LM parameters given by Eq.~(\ref{eq:gprime}), the critical density of $\pi^0$ condensation in symmetric nuclear matter and that of $\pi^c$ condensation in neutron matter have been estimated to be less than 2$\rho_0$ for both condensations.\cite{sst99} 
It is worth mentioning that the EOS of pion-condensed phase in neutron matter was obtained with relaxing the universality for the LM parameters by the parameter sets ($g'_{NN}$, $g'_{N\Delta}$)=(0.6$-$0.8,0.4$-$0.5) together with the assumption $g'_{N\Delta}$=$g'_{\Delta\Delta}$.\cite{mrt93} The resulting EOS lie in the intermediate region bounded by the previous two curves based on the universality and smaller values of $g'=0.5-0.6$. 

\subsection{$S$-wave kaon condensation}
\label{subsec:kaon}

The driving force for kaon condensation is the $s$-wave kaon-nucleon 
($K N$) interactions, which consist of the scalar and vector interactions.\cite{kn86} The former is 
simulated by the $KN$ sigma term $\displaystyle\Sigma_{KN}$ [=$\frac{1}{2}(m_u+m_s)\langle N|\bar uu+\bar ss | N\rangle$ ], which is larger than the $\pi N$ sigma term by an order of magnitude due to the large strangeness quark mass $m_s$, reflecting the violation of chiral symmetry in SU(3) case. The vector interaction is proportional to the nucleon $V$-spin density, which is written as $\displaystyle\psi^\dagger \frac{1}{2}(V_3+\sqrt{3}V_8)\psi=
\frac{1}{4}\psi^\dagger (\tau_3+3)\psi$ with $V_a$ being the vector charge. 
The $V$-spin of the proton (the neutron) is 1 (1/2), so that the antikaon ($K^-$) feels attractive from both proton and neutron in the vector channel, while $K^+$ feels repulsive. 

Empirically, the $s$-wave $KN$ scattering lengths are obtained as 
$a(K^+p)$=$-$0.33 fm, $a(K^+n)$=$-$0.16 fm, $a(K^-p)$=($-$0.67 +$i$ 0.64) fm, $a(K^-n)$=(0.37 +$i$ 0.60) fm.\cite{m81} The $K^- n$ interaction is attractive. The repulsive $K^-p$ scattering length is due to tha pole contribution of the $\Lambda$(1405) which lies below the $KN$ threshold by 30 MeV. 

 In neutron-star matter, the lowest excitation energy $\omega_{\rm min}$ 
of the $K^-$ decreases with increase in baryon number density $\rho_{\rm B}$ 
due to the $s$-wave $KN$ attractions. At a critical density,  
$\omega_{\rm min}$ becomes equal to the kaon chemical potential 
$\mu_K$\footnote{
In a chemically equilibrated matter, the kaon chemical potential is equal to the electron chemical potential $\mu_e$. In this paper, we put $\mu\equiv\mu_K=\mu_e$, and call it the charge chemical potential.} 
As a result, the distribution function $f_K({\bf p}_K)$ for the $K^-$ diverges at $\omega=\mu$, and the $K^-$ appears macroscopically as a Bose-Einstein 
condensate (BEC).\cite{mt92,bkrt92} 
The critical density for $K^-$ condensation, $\rho_{\rm B}^{\rm C}(K^-)$, is estimated as 
$\rho_{\rm B}^{\rm C}(K^-) =3-4\rho_0$ depending on the value of $\Sigma_{KN}$. 

Formation of kaon condensates is mainly controlled by {\it weak} interactions, in particular, by the thermal kaon process, $nn\rightarrow npK^-$, which gives the most efficient reaction rate among the nonequilibrium weak reaction processes at high temperature and density relevant to protoneutron stars. 
Kinetics of kaon condensation in newly born neutron stars has been investigated through obtaining time scales of saturation of thermal kaons, growth and subsequent saturation of kaon condensates and relaxation to chemically equilibrated state.\cite{mti00} It has been found that the time scale of the development of kaon condensates is much smaller than that of the cooling (several tens of seconds) for the relevant temperatures of protoneutron stars. The time delay for the formation of kaon condensates may not have much influence on the delayed collapse of a neutron star. 

Recently, some authors examined the possibility of kaon 
condensation in neutron stars by taking into account many-body 
effects such as the Pauli-blocking and the nucleon-nucleon 
correlations.\cite{wrw97,chp00} 

\subsection{Kaon dynamics in nuclear experiments}
\label{subsec:kaon-dynamics}

Motivated by studies of kaon condensation, the in-medium kaon properties have been investigated from both theoretical and experimental viewpoints. The off-shell properties of the kaon self-energy and its model dependence have been discussed.\cite{lbmr95,ynmk93,tw95} The kaon optical potential in nuclear matter has been deduced  by the use of chiral models,\cite{ro00} a G-matrix method with consistently taking into account the modification of kaons in medium,\cite{trp02} dispersion relation approach\cite{sc98} and so on. Several authors obtained the $K^-$ optical potential phenomenologically from kaonic atom data.\cite{fgb94,hot00,cfgm01} 
There is still controversy on the strength of the kaon optical potential, although recent 
experimental results on the subthreshold $K^+K^-$ production in 
relativistic heavy-ion collisions and  proton-nucleus collisions 
suggest a substantial decrease in the antikaon effective mass.\cite{s99} 

It should be noted that deeply bound kaonic nuclei have been recently 
proposed based on the strongly attractive kaonic potential.\cite{ay02,k99}  In Ref.~\citen{ay02}, a very high-density state comparable to that of neutron stars is predicted to form inside the nuclei.  If experimental achievement of such nuclei is established, it will provide us with information on kaon dynamics in high-density matter at zero  temperature, especially, the possible existence of kaon condensation. 

\subsection{$P$-wave kaon condensation}
\label{subsec:p-wave-kaon}

\noindent\begin{figure}
\noindent\begin{minipage}[l]{0.5\textwidth} 
\epsfxsize=\textwidth
\centerline{\epsfbox{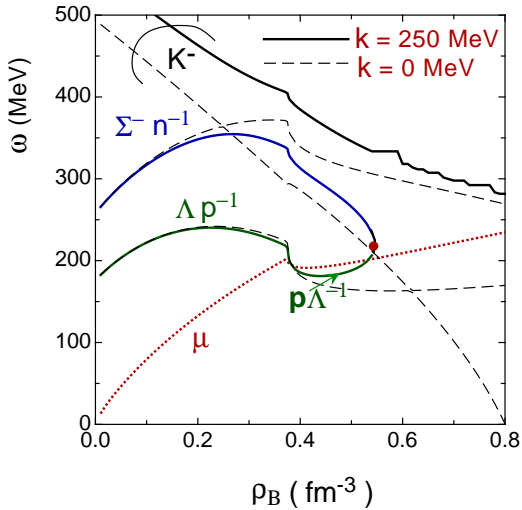}}
\caption{The excitation energy $\omega$ as functions of the baryon number density $\rho_{\rm B}$. }
\label{fig2}
\end{minipage}~
\begin{minipage}[r]{0.50\textwidth}
\ \ Recently, mixing of hyperons ($Y$) such as $\Lambda$, $\Sigma^-$, $\Xi^-$,  $\cdots$ as well as neutrons, protons, and leptons in neutron-star matter has been widely discussed motivated by  hypernuclear experiments.\cite{ekp95}  The interplay between kaon condensation and hyperons has been studied by the use of the SU(3)$_{\rm L}\times $SU(3)$_{\rm R}$ chiral effective Lagrangian where both $p$-wave $KNY$ and $s$-wave kaon-baryon interactions are incorporated.\cite{m02} The baryonic potentials in hyperonic matter deduced from the baryon-baryon interactions have also been taken into account. 

\ \ In Fig.~\ref{fig2}, the kaon minimal excitation energies $\omega$  are shown as functions of baryon number density $\rho_{\rm B}$ at a typical kaon momentum $|{\bf k}|$ and for the $Kn$ sigma term $\Sigma_{Kn}\simeq $ 300 MeV.  In this calculation, 
\end{minipage}
\end{figure}
\vspace{-0.55cm}

\noindent only the $\Lambda$ and $\Sigma^-$ have been taken into account as hyperons for simplicity. There appear collective $\Sigma^-$ particle-neutron hole mode (denoted as $\Sigma^- n^{-1}$) having the $K^-$ quantum number and proton particle-$\Lambda$ hole mode  (denoted as $p\Lambda^{-1}$) having the $K^+$ quantum number. 
One can see that the $\Sigma^- n^{-1}$ and 
$p\Lambda^{-1}$ modes merge at certain density ($\rho_{\rm B}\simeq$0.55 fm$^{-3}$). At this density and momentum, the double-pole condition for the kaon inverse propagator $D_K^{-1}$, $D^{-1}_K (\omega, {\bf k}; \rho_{\rm B})=0$, $\partial D^{-1}_K( \omega, {\bf k}; \rho_{\rm B})/\partial \omega=0$ is satisfied, and the system is unstable with respect to a pair creation of [$\Sigma^- n^{-1}$] and [$p\Lambda^{-1}$] modes. This instability originates from the $p$-wave kaon-baryon interaction and its mechanism is similar to that of pion condensation. 

In the $p$-wave kaon-condensed phase, baryons exist as quasiparticles, $|\tilde p_s\rangle$=$\alpha |p_s\rangle+\beta|\Lambda_s\rangle+\gamma|\Sigma^0_s\rangle$, $|\tilde n_s\rangle$=$\delta |n_s\rangle+\epsilon|\Sigma^-_s\rangle$, etc. with the spin state $s$(=$\pm$1/2).  

\section{Rapid cooling processes in meson condensations}

\subsection{Neutrino emission in meson condensates}

In ordinary neutron-star matter without exotic states, the nucleons and electrons are strongly degenerate, and the energy-momentum conservation is not satisfied for the following neutrino and antineutrino emission processes as one-nucleon processes, $n\rightarrow p+e^-+\bar\nu_e$, $p+e^-\rightarrow n+\nu_e$, which is called the direct Urca process\footnote{
The possibility of the direct Urca process depends on the density-dependence of the nuclear symmetry energy.\cite{lpph91,fmtt94} }. Therefore, a spectator nucleon participates in the reaction so as to meet the kinematical condition, and the two-nucleon process, $ n+n\rightarrow n+p+e^-+\bar\nu_e$, $n+p+e^-\rightarrow n+n+\nu_e$, becomes possible. This reaction is the standard cooling process in the normal phase and is called the modified Urca process.\cite{fm79}
\begin{figure}
\noindent\begin{minipage}[l]{0.5\textwidth} 
\epsfxsize=\textwidth
\centerline{\epsfbox{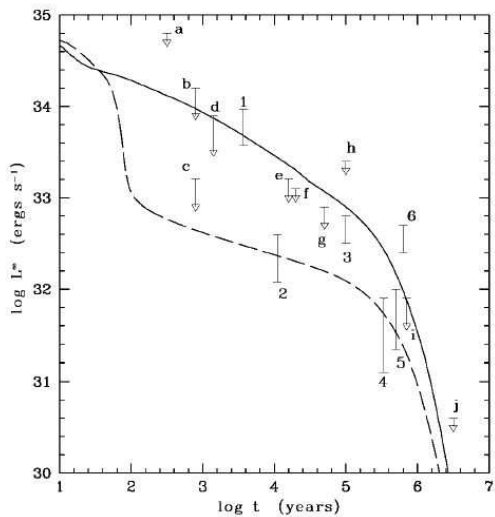}}
\caption{The photon luminosities as functions of the age of a neutron star, taken from Tsuruta et al., Ref.~44). 
The solid line denotes the standard cooling without any exotic agents and the dashed line including the extra rapid cooling under pion condensation. }
\label{fig3}
\end{minipage}~
\begin{minipage}[r]{0.50\textwidth}
\ \ In Fig.~\ref{fig3}, we show a standard cooling curve of neutron stars without any exotic phase with the gravitational mass $M$=1.2$M_\odot$ ($M_\odot$ being the solar mass) by a solid line,  which is taken from Tsuruta et al.\cite{ttt02} The photon luminosity observed at infinity, which is connected with the surface temperature of a neutron star, is given as a function of the age of a neutron star.  It should be noted that there are several point sources observed, (c): PSR J0205+6449 in 3C58, (2): the Vela pulsar, (4) Geminga of which the surface temperatures are too low to be explained by any  standard cooling theories. Extra cooling mechanisms leading to rapid cooling of neutron stars are needed for some observed point sources. 
\end{minipage}
\end{figure}

In the presence of meson condensates, the classical meson field supplies the system with energy and momentum, and the one-nucleon process becomes possible. 
The following reactions have been mainly considered. One is the quasiparticle ($\eta$) Urca process, 
\begin{equation}
\eta \ + \ \langle \pi^- \rangle\rightarrow 
\eta \ + \ e^- \ + \ \bar\nu_e \ , \ \ 
\eta \ + \ e^- \ \rightarrow \eta \ + \ 
\langle \pi^- \rangle \ + \nu_e \ , 
\label{eq:pion-urca}
\end{equation}
for pion condensation,\cite{m77} and another is the 
 Kaon-induced Urca (KU) process, 
\begin{equation}
n \ (p) \ + \ \langle K^- \rangle\rightarrow 
n \ (p) \ + \ e^- \ + \ \bar\nu_e \ , \ \ 
n \ (p) \ + \ e^- \ \rightarrow n \ (p) \ + \ 
\langle K^- \rangle\ + \nu_e \ , 
\label{eq:kaon-urca}
\end{equation} 
for the $s$-wave kaon condensation.\cite{bk88,fmtt94}

\begin{figure}
\noindent\begin{minipage}[l]{0.5\textwidth} 
\epsfxsize=\textwidth
\centerline{\epsfbox{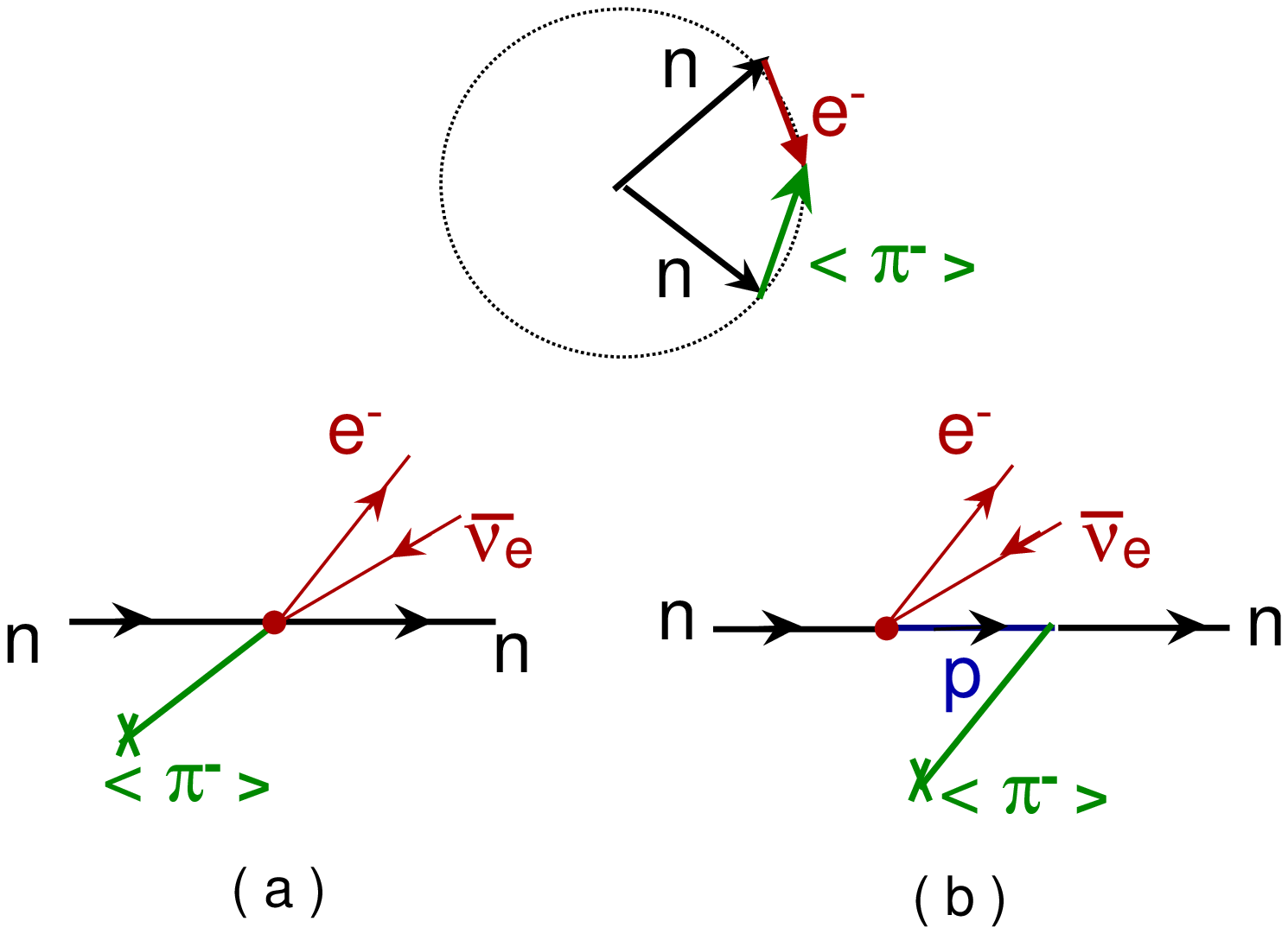}}
\caption{The upper part representing the momentum conservation between the participating degenerate nucleons, leptons, and the classical $\pi^c$ field. The lower part showing the lowest order diagrams for the quasiparticle Urca process in charged pion condensates.  }
\label{fig4}
\end{minipage}~
\begin{minipage}[r]{0.50\textwidth}
\epsfxsize=0.5\textwidth
\centerline{\epsfbox{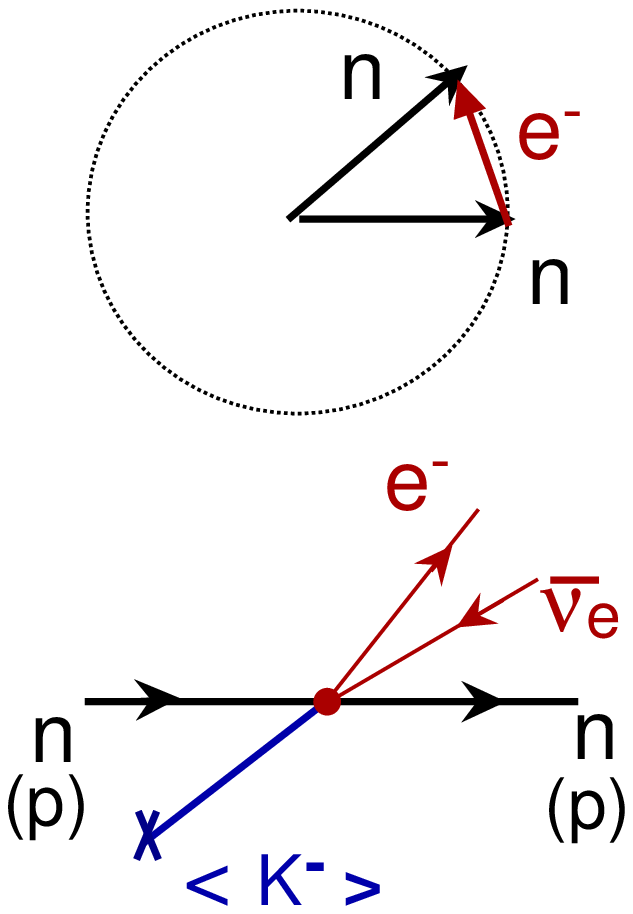}}
\caption{The same as Fig.~4 but for the Kaon-induced Urca process in the $s$-wave kaon condensates. }
\label{fig5}
\end{minipage}
\end{figure}

These weak processes can be treated in the framework of chiral symmetry.\footnote{
Neutrino absorption processes in meson condensates, such as $\nu_e+\eta+\langle\pi^-\rangle\rightarrow e^-+\eta$  and $\nu_e +n(p)+\langle K^-\rangle\rightarrow e^-+n(p)$, which are relevant to neutrino opacities, can also be considered within this framework. \cite{ss77,mti03}}  In Fig.~\ref{fig4}, we show the relation between the baryon and lepton momenta on the Fermi surface contributing to  momentum conservation, and the lowest order diagrams for the quasiparticle Urca process in charged pion condensates.  Fig.~\ref{fig5} is the same but for the $s$-wave $K^-$ condensation. 

\noindent

Here we take up only the KU process associated with the $s$-wave $K^-$ condensation. 
The relevant weak Hamiltonian for KU is given by the current$\times$current type, 
$\displaystyle H_{\rm W}={G_F\over \sqrt 2}
 J_{\rm h}^\mu\cdot l_\mu+{\rm h.c.}$ with $l_\mu$ [=$\bar\psi_e \gamma_\mu (1-\gamma_5)\psi_\nu $  ] being the charged leptonic current and the charged hadronic current,  
$J_{\rm h}^\mu=\cos\theta_{\rm C} (V_{1+i2}^\mu -A_{1+i2}^\mu)
+\sin\theta_{\rm C} (V_{4+i5}^\mu -A_{4+i5}^\mu) $, 
where $\theta_{\rm C}(\simeq 0.24)$ is the Cabibbo angle,  
and $V_a^\mu$ and $A_a^\mu$ are the vector and axial-vector currents, 
respectively. 
The matrix elements are given 
 from the transformed Hamiltonian: 
\begin{equation}
 {\widetilde H}_{\rm W}=\hat U_K^{-1}H_{\rm W}\hat U_K 
={G_{F}\over \sqrt 2}
{\widetilde J}_{\rm h}^\mu l_\mu
+{\rm h.c.} \ ,
\label{th}
\end{equation}
where $\hat U_K=\exp(i\mu_K Q)\exp(i\theta Q_4^5)$, which is a unitary operator generating a kaon-condensed state, $| K\rangle=\hat U_K |0\rangle$. By the use of current algebra, the tranformed hadronic current is shown to have a form, 
\begin{eqnarray}
 {\widetilde J}^\mu_{\rm h}
&=&\hat U_K^{-1}J^\mu_{\rm h}\hat U_K \cr
&=&e^{-i\mu_K t}\bigg\lbrack\cos\theta_c\bigg\{
(V_{1+i2}^\mu-A_{1+i2}^\mu)\cos(\theta/2)
+i(V_{6-i7}^\mu-A_{6-i7}^\mu)\sin(\theta/2)\bigg\} \cr
&+&\sin\theta_c\bigg\{(V_4^\mu-A_4^\mu)+i\cos\theta(V_5^\mu-A_5^\mu)
-i\sin\theta\Big(V_V^\mu-A_V^\mu\Big)\bigg\}\bigg\rbrack \ , 
\label{eq:tjh}
\end{eqnarray} 
where $V_V^\mu\equiv(V_3^\mu+\sqrt{3}V_8^\mu)/2$ and $A_V^\mu\equiv(A_3^\mu+\sqrt{3}A_8^\mu)/2$. 
The matrix elements for KU come from the term proportional to $\sin\theta$ in the last term of Eq.~(\ref{eq:tjh}). One can see that the kaon chemical potential $\mu_K$ is supplied to energy conservation of the reaction, stemming from the time dependent phase factor 
$\exp(-i\mu_Kt)$. 

The weak reactions (\ref{eq:pion-urca}) and (\ref{eq:kaon-urca}) give large neutrino emissivities since the available phase space is large as compared with that of two-nucleon reactions like the modified Urca process. As an example, the ratio of the emissivity for KU to that for the modified Urca process is roughly $O(\mu_{\rm B}/T)^2\sim 10^{6-8}$ with $\mu_{\rm B}$ being the nucleon chemical potential depending on the temperature. 
Once such an exotic phase appears,  it has a significant effect on thermal evolution of neutron stars.\cite{u94,t98} 

In case of $p$-wave kaon-condensation, enhanced cooling will occur through the 
weak processes, $\tilde Y +\langle K^-\rangle\rightarrow  \tilde Y +e^-+\bar\nu_e$, 
$\tilde Y+ e^-\rightarrow \tilde Y +\langle K^-\rangle +\nu_e$ (for $\tilde Y$=$\tilde p$, 
$\tilde\Lambda$, $\tilde n$, $\tilde \Sigma^-$), where $\tilde Y$ is a quasiparticle whose state is given by superposition of the nucleon and hyperon states.  
These weak processes may also lead to the increase of the absolute value of total 
negative strangeness of the system. 

\subsection{Effects of superfluidity}
\label{subsec:super}

The neutrino emissivities themselves in meson condensates lead to too rapid cooling of the neutron stars that need an enhanced cooling scenario, as other cooling agents such as quarks,  hyperons, and the direct Urca process do.\cite{t98} Thus suppression mechanisms are needed in order to counterbalance the too rapid cooling.  Nucleon superfluidity suppresses both specific heat of nuclear matter and neutrino emissivities owing to the existence of an energy gap $\Delta$. For the neutrino emissivities, the available phase space is largely restricted due to the energy gap $\Delta$. 
 
 In an inner core of a neutron star, the $^3$P$_2$ pairing is most favorable for the neutrons, 
 while $^1$S$_0$ pairing is for the protons, which is deduced from the nucleon-nucleon scattering phase shifts for the isospin $T=1$ pair.\cite{dj03} 
For the normal phase, Takatsuka and Tamagaki studied the neutron $^3$P$_2$ (+$^3$F$_2$) pairing by generalizing the  Bogoliubov transformation to the pairings with nonzero angular momentum.\cite{t72} 

They also studied pairing of quasineutrons ($\eta$) under pion condensation.\cite{tt82} They found that the energy gap is sensitive to the effective nucleon mass $m_\eta^\ast$ at a high density. 
They obtained $m_\eta^\ast (\pi^c) \simeq m_n^\ast ({\rm Normal})$=(0.8$-$0.7 )$m_N$, and $m_\eta^\ast (\pi^0{\rm -}\pi^c) \simeq m_n^\ast(\pi^0)\sim $0.9 $m_N$ for $\rho_{\rm B}=1-3 \rho_0$ with $m_N$ being the free nucleon mass. Note that the effective masses in the $\pi^0$-condensed and combined $\pi^0$-$\pi^c$ condensed  phases do not decrease much even at high densities $\sim 3 \rho_0$, reflecting the localization picture of the ALS structure. The large effective mass results in large pairing energy gap. Thus the nucleon superfluid in the $^3$P$_2$ (+$^3$F$_2$) channel can exist at rather high densities with the existence of $\pi^0$ condensation. 
On the other hand, the quasineutron pairing in the $\pi^c$-condensed phase has been shown to be less favorable owing to the attenuation effect coming from a factor $1-\cos\phi\cdot\sin\phi$ with the mixing angle $\phi$. 

With the effects of quasineutron superfluidity in $\pi^0$-$\pi^c$ condensed phase in the inner core of a neutron star, the cooling behavior of the pion-condensed neutron star has been obtained as shown by a dashed line in Fig.~\ref{fig3}. It has been shown that the anomalously low temperatures for several neutron stars including the three candidates in this figure can be consistently explained with the rapid cooling under $\pi^c$ condensation and its moderate suppression due to the coexistent quasineutron superfluidity under $\pi^0$ condensation. 

\section{Static and dynamic properties of meson-condensed neutron stars}

The existence of meson condensations has important implications for static and dynamic properties of neutron stars. 

\subsection{Softening of the EOS}
 \label{subsec:eos}
 
A phase transition to a meson-condensed phase leads to softening of the EOS, making a neutron star compact with higher central density and smaller radius than normal neutron stars. As a result, a maximum gravitational mass of a neutron star becomes small.\cite{lp01} 
From the observation of PSR 1913+16, its mass was obtained as $ 1.442\pm 0.003 M_\odot$.\cite{tw89} The statistical analysis of mass determination of binary pulsar systems gave the upper limit of the mass $\sim 1.6 M_\odot$,\cite{f94} and measurements of masses of neutron stars in $X$-ray binaries led  to a narrow mass range with 1.35$\pm$0.27$M_\odot$.\cite{tc99} The mass should be even larger ($\sim$ 2.0$M_\odot$),  if the recent analyses from the observations of the quasiperiodic oscillations are confirmed.\cite{mlp98} The mass-radius relation was constrained from the analyses of the gravitationally redshifted absorption lines of the $X$-ray bursts.\cite{ft86} 
The mass inferred from the EOS should be consistent with these observations. 

The significant softening will make phase transition of a first order. Here we mention consequences from a first-order phase transition induced by kaon condensation, following recent development of studies of kaon condensation. 
 We can also see common properties in case of pion condensation.\cite{mttt93} 

 Effects of first-order phase transition on the internal structure of neutron stars have been discussed by the use of the Maxwell's construction by which the phase equilibrium was obtained. However, it has been pointed out by Glendenning that the Gibb's condition determines the appropriate phase equilibrium in case more than one chemical potentials are present corresponding to conserved charges such as the electromagnetic charge, baryon number and so on.\cite{g01} According to the Gibb's condition, the first-order phase transition may imply 
a mixed phase where droplets of a kaon condensate are immersed 
in the normal phase. Recently the effects of Coulomb energy and surface tension on the mixed phase have been discussed.\cite{hps93,cgs00,nr01,n02,vyt02} There are several works on a possible formation of the crystalline mixed phase on a Coulomb lattice and on the stability of the lattice.\cite{cgs00,z02}The nucleation of a kaon droplet has also been considered.\cite{n02} On the other hand, it has been shown that the Maxwell's construction is substantially good in case of the large Coulomb screening effects.\cite{vyt02} 

The first-order phase transition has also important effects on the dynamical properties of neutron stars. A minicollapse from a supercompressed metastable neutron star to a stable kaon-condensed star has been considered within a relativistic mean field model.\cite{fmmt96} The energy release after the minicollapse has been estimated, and it has been found that the released energy, $10^{49}-10^{52}$ erg depending on a scale of the first-order phase transition, is sufficient for explaining the observed anomalous gamma-ray burst energies. 

In relation to an early stage of neutron stars, a formation scenario of a low-mass black hole by a delayed collapse of a hot neutron star after a  phase transition to a kaon-condensed star has been presented by Brown and Bethe,\cite{bb94} and a numerical simulation has been elaborated.\cite{bst95} 
Delayed collapse of protoneutron stars has been discussed in detail by several authors.\cite{p00,ty99} Tatsumi and Yasuhira obtained the EOS with kaon condensates at finite temperature including thermal effects and neutrino degeneracy. Based on this EOS, they discussed the gravitational stability of the protoneutron stars against a phase transition to a kaon-condensed neutron star during the deleptonization era.\cite{ty99}

\subsection{Pulsar glitches}
\label{subsec:glitch}

Pulsar glitches give information on the inner structure of neutron stars. They are characterized by sudden discontinuous spin up of rotation and its subsequent relaxation to a steady state of radio pulsars. From the temporal behavior of the angular velocity  after the spin up, one can see two kinds of relaxation time scales, one is long with $\tau_1\sim$ 80 days and the other is short with $\tau_2\sim$ several days. According to the usual glitch scenario,  sudden spin up is attributed to a catastrophic unpinning of vortices of the nucleon $^1$S$_0$ superfluids from nuclei in the inner crust region of a neutron star and the gradual decrease of rotation results from creeping of vortex toward the outside region of the star.\cite{a84,m95} 
The vortex creep theory has been considered as one of the standard postglitch models.\cite{a84}

On the other hand, Takatsuka and Tamagaki presented a corequake model with combined $\pi^0$-$\pi^c$ condensation for giant glitches like the Vela pulsar.\cite{tt88} This model is based on the $\pi^0$-$\pi^c$ condensed phase in a central core region of a neutron star. The ALS structure accompanying $\pi^0$ condensation presents a solid core. As the pulsar spins down, stress of the core is accumulated, and finally the core cracks. At that time, the pulsar suddenly spins up as a result of angular momentum conservation.  Large energy release due to the 
corequake is carried away through a rapid cooling by $\pi^c$ condensation before the next glitch. The two time scales $\tau_1$ and $\tau_2$ which are observed in the postglitch timing behavior can be explained by the two relaxation times of the $^1$S$_0$ superfluids in the crust and the $^3$P$_2$ superfluids in the core. They also succeeded in descriminating giant glitches like the Vela pulsar and small ones (smaller ratio of change of angular velocity to stationary angular velocity than that of giant glitches by two-orders of magnitude) like the Crab pulsar: The Vela
pulsar is assumed to have a large mass $\sim 1.6M_\odot$. Then the central density in the inner core exceeds the critical density of $\pi^0$-$\pi^c$ condensation, so that the core has the $\pi^0$-$\pi^c$-condensed phase, which induces the giant glitch by a corequake. On the other hand, the Crab pulsar is assumed to have a smaller mass $\sim 1.2 M_\odot$, where the central density does not reach the critical density of the $\pi^0$-$\pi^c$ condensation, so that there is no ALS solid core. In this case, only a crustquake, which is smaller in magnitude than the giant glitches, occurs. 

Thus the new corequake model based on pion condensation has a possibility of consistently explaining the observed glitch phenomena. 

\section{Coexistence of pion and kaon condensations}
\label{sec:pi-k}

According to the preceding results, pion condensation appears at a lower baryon number density ($\sim 2\rho_0$) as compared with that for kaon condensation ($\sim 3-4 \rho_0$). 
Possible coexistence of pion and kaon condensations ($\pi$-$K$ condensation) in neutron-star matter has been studied in a simple model based on the SU(3)$_{\rm L}\times $SU(3)$_{\rm R}$ chiral effective Lagrangian.\cite{kn86} 
The classical meson fields in the nonlinear representation are taken to be plane-wave forms:  
\begin{equation}
\pi^-({\bf r}, t)=\frac{f}{\sqrt{2}}\theta\cdot\cos\beta e^{-i(\mu t-{\bf p}_\pi\cdot{\bf r})}  \ \ , \ 
K^-({\bf r}, t)=\frac{f}{\sqrt{2}}\theta\cdot\sin\beta e^{-i(\mu t-{\bf p}_K\cdot{\bf r})}
\label{eq:pi-k}
\end{equation}
with a parameter $\beta$ representing the relative strengh of pion and kaon condensations. 
\vspace{-0.5cm}

\noindent\begin{figure}
\noindent\begin{minipage}[l]{0.5\textwidth} 
\epsfxsize=\textwidth
\centerline{\epsfbox{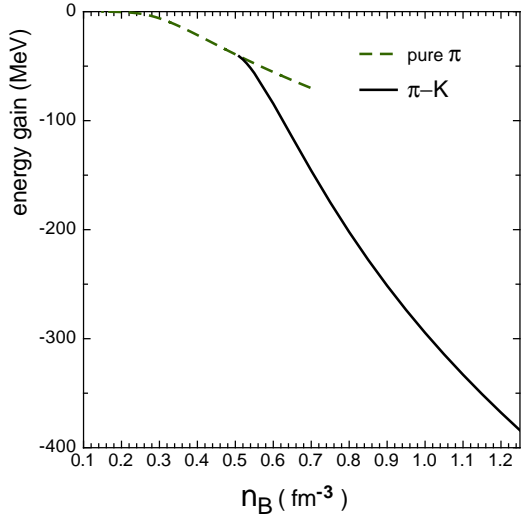}}
\caption{The energy difference per baryon between the $\pi$-$K$ condensed phase and the normal neutron-star matter as a function of the baryon number density.}
\label{fig6}
\end{minipage}~
\begin{minipage}[r]{0.50\textwidth}
\ \ One had a picture that charged pion condensation is suppressed as kaon condensation grows with increase in baryon number density. However,  there was a problem in the original works that the pion spatial momentum diverges beyond a scope of chiral perturbation theory at a density where pion condensation disappears in a kaon-condensed phase. We have reexamined $\pi$-$K$ condensation by taking into account the recoil terms between mesons and baryons in the nonrelativistic expansion of the baryonic Hamiltonian with respect to 1/$m_N$ 
in order to avoid the divergence of the pion momentum. 
Only the charged pion and kaon have been considered as the octet mesons. 
\end{minipage}
\end{figure}
\vspace{-0.5cm}

In Fig.~\ref{fig6}, we show the energy difference per baryon between the $\pi$-$K$ condensed phase and the normal neutron-star matter as a function of the baryon number density by a solid line. The dashed line is for the pure $\pi^c$-condensed case.  One can see that kaon condensates begin to appear in the $\pi^c$ condensed phase at the baryon number density $\sim$0.5 fm$^{-3}$ and that the energy gain due to kaon condensates is added to that due to pion condensates as density increases. 

It has been concluded that pion and kaon condensations can coexist over the relevant densities in neutron stars, increasing their amplitudes monotonically with increase in density. In the ground state of the $\pi$-$K$ condensed phase, only lower energy eigenstate of quasiparticles, $|\eta\rangle=\alpha |n\rangle+\beta|p\rangle$ which tends to $|n\rangle$ in the limit $\theta\rightarrow 0$, fills the Fermi sea ("one-Fermi sea" approximation is valid), which is a different picture from a pure kaon-condensed case: In the $K^-$ condensed phase, both protons and neutrons fill independent Fermi seas.   
Toward realistic studies of $\pi$-$K$ condensation, one has to take into account the effecs of baryonic potentials in matter, 
short-range correlations between baryons, vertex renormalizations at the meson-baryon vertices, hyperon-mixing and other subthreshold resonances such as $\Lambda$(1405), $\Sigma$(1385), $\Delta$(1232) etc. 
 
\section{Meson condensation and color superconductivity}
\label{sec:color}

Condensation of the NG bosons is also found in quark matter. 
The NG boson masses in CFL phase are given according to the inverse mass splitting in the form, 
\begin{eqnarray}
m_{\pi^\pm}&=&\mp\frac{m_d^2-m_u^2}{2p_F}+\Big\lbrack\frac{4A}{f_\pi^2}(m_u+m_d)m_s\Big\rbrack^{1/2} \ , \cr
m_{K^\pm}&=&\mp\frac{m_s^2-m_u^2}{2p_F}+\Big\lbrack\frac{4A}{f_\pi^2}m_d(m_u+m_s)\Big\rbrack^{1/2}  \ , \cr
m_{K^0, \bar K^0}&=&\mp\frac{m_s^2-m_d^2}{2p_F}+\Big\lbrack\frac{4A}{f_\pi^2}m_u(m_d+m_s)\Big\rbrack^{1/2} \ , 
\end{eqnarray}
where $m_i$ ($i=u, d, s$) are the quark masses and $p_F$ the Fermi momentum.\cite{bs02} 
In the CFL phase, the mass of the strange quark $m_s$ works as the effective chemical potential $\mu_{\rm eff}=m_s^2/(2p_F)$. For a large $m_s$, the mass for $K^0$ or $K^+$ first vanishes, and the system becomes unstable with respect to creation of the kaonic mode $K^0$ ($\sim$[$\bar u \bar s$][$du$] ) or  $K^+ $($\sim$[$\bar d \bar s$][$du$] ). 

Elementary excitations from kaon condensates in the CFL phase have been considered, where  neutrino emissivities from decay and scattering of the NG bosons in the CFL phase have been obtained.\cite{jps02} 

There is a similarity between the "low density" kaon condensation in hadronic matter and the "high density" kaon condensation in the CFL matter: In both phases, the explicit chiral symmetry breaking triggers the onset of kaon condensation. It is an open problem whether there is a continuity between these meson condensations in hadronic matter and quark matter. 

\section{Summary and concluding remarks}

We have reviewed the present studies of meson condensations as a typical hadronic phase in high-density matter. One can find common features between meson-condensed phase in hadronic matter and quark phase through various astrophysical phenomena: 
(1) Meson-condensed phase leads to softening of the EOS of hadronic matter, affecting static and dynamic properties of neutron stars, as is the case with quark phase. (2) The ALS structure accompanying $\pi^0$ condensation represents solid-like aspects. The existence of the quantum solid is essential to explain the giant glitches of pulsars in a corequake model. On the other hand, a quark alternating-layer-spin-flavor (ALSF) model has been presented as an intermediate phase between hadronic phase and quark phase, where quarks in the perturbative vacuum have a layered structure with a specific spin-flavor order and the $\pi^0$ condensed field in the nonperturbative vacuum persists in the region between the adjacent layers.\cite{qtt93} (3) The crystalline mixed phase of kaon condensation on a Coulomb lattice may have an influence on pinning of nucleon superfluid vortices formed in neutron star core region, which may have a relevance to a vortex creep model for pulsar glitches. In quark phase, there is a crystalline color superconductivity similar to the LOFF state,\cite{loff64} where pairing quarks have nonzero momentum, leading to gaps with a periodic ordering of a crystalline structure.\cite{abr01} (4) Neutrino emission processes are enhanced in meson condensates, which leads to rapid cooling of neutron stars. Such exstra weak processes exist also in quark phase,\cite{i82} whereas there is much suppression of neutrino emissivity in case quarks are in color superconductivity due to a large energy gap of order 50$-$100 MeV.\cite{a03} Throughout these features (1)$-$(4), pairing of Fermions, nucleon and/or quasinucleon superfluidity in hadronic phase and color superconductivity for quark phase, provides important aspects for a consistent explanation of the related phenomena, glitches and rapid cooling of neutron stars. 

It is interesting to clarify a close connection between hadronic phase and quark phase beyond the apparent phenomenological similarities listed above. The basic features inherent in QCD, chiral symmetry and quark confinement/deconfinement,  play an important role in connecting hadron dynamics and quark-gluon dynamics,  leading to a unified description of the QCD phase diagram. 
A theoretical approach from the lattice QCD will give complementary understanding of the QCD phase diagram. Observations of compact stars by Chandra, XMM/Newton, HST etc. and experiments planned at JHF and GSI are also promising. 

\section*{Acknowledgments}
The author is grateful to the organizers of the conference "Finite Density QCD at Nara" 
for giving him an opportunity to review present studies of meson condensations in hadronic matter. Part of this work is based on recent collaboration with T. Tatsumi, to whom I would like to thank for discussion. This work is supported in part by the Japanese Grant-in-Aid for 
Scientific Research (C) of the Ministry of Education, Culture, Sports, Science, 
and Technology (Nos. 12640289). 

\end{document}